\documentclass[manuscript,screens]{acmart}
\AtBeginDocument{%
  \providecommand\BibTeX{{%
    \normalfont B\kern-0.5em{\scshape i\kern-0.25em b}\kern-0.8em\TeX}}}

\setcopyright{acmcopyright}
\copyrightyear{2023}
\acmYear{2023}
\acmDOI{XXXXXXX.XXXXXXX}

\acmConference[GoodIT '23]{ACM International Conference on Information Technology for Social Good}{September 06--08, 2023}{Lisbon, Portugal}
\acmBooktitle{GoodIT '23: ACM International Conference on Information Technology for Social Good, 6-8 September, 2023 Lisbon, Portugal} 
\acmPrice{15.00}
\acmISBN{978-1-4503-XXXX-X/18/06}

\begin{document}

\title{Data Discovery for the SDGs: A Systematic Rule-based Approach}

\author{Yuwei Jiang}
\orcid{0009-0009-0999-5378}
\author{David Johnson}
\email{david.johnson@im.uu.se}
\orcid{0000-0002-2323-6847}
\affiliation{%
  \institution{Department of Informatics and Media, Uppsala University}
  \streetaddress{Box 513}
  \city{Uppsala}
  \country{Sweden}
  \postcode{751 20}
}
\renewcommand{\shortauthors}{Jiang and Johnson}

\begin{abstract}
In 2015, the United Nations put forward 17 Sustainable Development Goals (SDGs) to be achieved by 2030, where data has been promoted as a focus to innovating sustainable development and as a means to measuring progress towards achieving the SDGs. In this study, we propose a systematic approach towards discovering data types and sources that can be used for SDG research. The proposed method integrates a systematic mapping approach using manual qualitative coding over a corpus of SDG-related research literature followed by an automated process that applies rules to perform data entity extraction computationally. This approach is exemplified by an analysis of literature relating to SDG 7, the results of which are also presented in this paper. The paper concludes with a discussion of the approach and suggests future work to extend the method with more advance NLP and machine learning techniques.
\end{abstract}

\keywords{knowledge discovery, data use, sustainable development, SDG, systematic mapping, named entity extraction}

\maketitle

\section{Introduction}
In 2015, the UN adopted a new agenda for Sustainable Development, including 17 Sustainable Development Goals (SDGs) and 169 related targets to be achieved by each member state by 2030 \citep{2030Agenda}. To monitor the achievement of SDGs, it is significant to obtain data and then conduct an analysis. Surveys and interviews are typical data collection methods in sustainable development. The advantage of surveys is that it is possible to solicit a large amount of data from target populations, for example, via a census. Interviews give researchers more in-depth information and allow for discussions between interviewers and interviewees to dig deeper into specific issues. However, they are time-consuming and cannot be generalised, especially on a large scale \citep{queiros2017strengths}. Therefore, the typical limitation of these traditional data collection methods is that they are often expensive and time-consuming \citep{escap2017innovative}.

With the rapid development of the Internet and technology, new data sources have emerged, such as data from social media and data gathered through different types of sensors. There are also various data-sharing resources, such as open databases to which the public has free access. Such online resources provide us with a new perspective on how we can monitor the progress of the SDGs. The diversity of data sources and recent advances in technologies like the Cloud and the Internet of Things (IoT) has made the speed of data generation reach an unprecedented level \citep{cai2015challenges}. We are currently in the Big Data era \citep{mcafee2012big}, where abundance of data bring opportunities to many fields, including sustainable development. Furthermore, data-driven technology such as artificial intelligence (AI) can contribute to the SDGs by enabling more accurate predictions, efficient resource allocation, and targeted interventions for achieving sustainable development goals \citep{vinuesa2020role}. While the traditional methods of data collection, such as household surveys, play a vital role in monitoring the development of society from a long-term perspective, other kinds of data may supplement traditional methods to provide real-time statistics \citep{wahlisch2020big}. In some cases, data can be used as proxies that are better than traditional methods for gathering information relating to sustainable development. For example, when natural disasters like earthquakes occur, emergency services may need to make decisions based on real-time data, such as the number of casualties and estimated economic loss; something is evident that surveys and interviews cannot provide in real-time \citep{mendoza2019nowcasting}.

There is a wide range of research exploring how to use various data sources to monitor the progress of the SDGs. Many different types of data include satellite imagery, sensing data, health records, social media data, and more \citep{macfeely2019big}. Open databases such as UNdata and World Bank Open Data also store and share many SDG-related data. A report commissioned by the UN Economic and Social Commission for Asia and the Pacific (ESCAP) systematically analysed and presented around 140 projects to show what types of data and innovative data-driven approaches could support tracking the fulfilment of SDGs \cite{escap2017innovative}. However, there are no dedicated resources, especially for those stakeholders in the SDGs, which describe what data sources may be helpful for monitoring or addressing the SDGs, except for occasionally published analyses such as those mentioned earlier presented by UN ESCAP. We, therefore, come to ask the following research question:

\begin{quote}
"What methodological approach can enable the discovery and analysis of data relevant to the SDGs?"
\end{quote}

In the remainder of this paper, we describe related work reported in the literature about infrastructures enabling data discovery and their relation to the SDGs and systematic approaches to knowledge discovery. We then follow with a description of our methodology for data source discovery. We then describe in the results section the application of our method to discovering data sources relevant to SDG 7. Finally, we discuss the implications for our approach and future lines of inquiry.

\section{Related work}

\subsection{Data infrastructures and the SDGs}
Data infrastructures are defined by \cite{kitchin2014data} as "...the institutional, physical and digital means for storing, sharing and consuming data across networked technologies". Gray et al. \cite{DataInfraLiteracy} state that data infrastructures contain data sets and describe how data could be used and shared. "Raw" data makes no sense unless people understand the context of the data, which is how it is generated and what it can be used for. Gray et al state that data infrastructures were not only data resources but were interpreted as a set of relations in terms of datasets, standards, policies, and more. Kitchin \cite{kitchin2014towards} points out that data infrastructures can be likened to recipes, where chefs describe how to use the ingredients. When considering these perspectives, we can think of data infrastructures as a means to make data available to others interested in it, such as data analysts, data scientists and software developers, and thus, to those who are likely to make the best use of the shared data.

UN ESCAP's report describes how Big Data, IoT, and AI already have great power in their fields, and a combination of these three technologies would create more value, especially in the area of sustainable development \cite{escap2017innovative}. IoT serves to gather a large amount of data from loads of different sources apart from traditional methods, and the data of various types and sources are said to characterise Big Data approaches, which are analysed through AI methods to explore patterns and help with decision-making. MacFeely \cite{macfeely2019big} states that if such methods were correctly applied, they would reduce the time spent on calculating SDG indicators and improve the accuracy of results. Not only have scholars noted that different types of data, and their combination thereof, have the potential for monitoring SDGs, but the UN has also realised the opportunities brought by the widespread abundance of data. There are various generalised research data sharing platforms such as Harvard's Dataverse\citep{king2007introduction}, as well as platforms for specific domains such as COPO  for the life sciences \citep{shaw2020copo}, and data catalogues such as FAIRsharing \cite{sansone2019fairsharing} aim to facilitate finding relevant data within the data diaspora. Within sustainable development, the UN Global Platform was set up to support the worldwide collaboration for data sharing for the SDGs \citep{GlobalPlatform}.

While efforts such as the UN Global Platform aim to provide a central resource for data sharing relating to the SDGs, it is still challenging to discover the relevant data for use in SDG activities. Even with these available data infrastructures, the there is still a gap towards organisations using data analytics to contribute towards the SDGs \citep{el2021value}. This paper aims to provide an approach to discovering data used in SDG research.

\subsection{Systematic approaches to knowledge discovery}
While data infrastructures may provide facilities to find data sources, they often rely on explicit contributions from a community or dedicated individuals to build and maintain such resources. For example, the infrastructures mentioned earlier (Dataverse, COPO and FAIRsharing) depend on researchers' contributions to deposit or index their data, and the UN Global Platform relies on maintenance by its own staff. Research is often reported in the literature, published in journals and conferences, and indexed by bibliographic databases (e.g., PubMed, Web of Science, JSTOR). Such publications report the research outputs rather than the specifics of the data they may have used. Thus, some work is required for data infrastructures to be up-to-date and curate themselves to report on new data types and sources (for brevity, we will refer to text mentioning data types and sources as "data entities").

Automated data discovery has frequently been proposed in the literature \citep{michener2006meta, fernandez2018aurum, farber2021datahunter}, but its adoption in scientific research has yet to be established. There exist general-purpose search engines (e.g. Google, Bing), knowledge engines (e.g., Wolfram Alpha, Watson) and, more recently, large language model-based tools (e.g., ChatGPT, Bard) that can assist in data discovery. There also exists only one dedicated dataset search engine, Google's Dataset Search. While these are powerful tools, they are also opaque in how they function and are designed to be general-purpose. When considering specific tasks within a specific domain, such as discovering data for particular SDGs, we often prefer a systematic and transparent approach to maintain trust in a given discovery. Thus, systematic research approaches remain relevant despite these powerful automated tools being available.

The Systematic Literature Review (SLR) is a method that typically aims to identify, evaluate, and synthesise a corpus research literature in order to address a research question that may or may not be related to the aims of the constituent research reported \citep{webster2002analyzing}. SLRs are characterised by defining a reproducible method of selecting, appraising, and combining research evidence relevant to that question. This method aims to minimise bias by performing an exhaustive search and quality appraisal of the literature and by making explicit the methods used for data extraction and synthesis.

The Systematic Quantitative Literature Review (SQLR) is similar to the SLR in its approach to providing a reproducible method while being comprehensive in review and minimising bias. However, SQLRs, as the name suggests, are quantitative, aiming to quantitatively summarise and analyse the literature \citep{pickering2014benefits}. This is through counting features of interest in the literature, such as the categories of studies published, geographical and temporal trends, types of methods used and characteristics of the results. In contrast to traditional SLRs, SQLRs do not aim to synthesise the published results but rather synthesise the studies' characteristics.

Similar to SQLRs, but differing in aims, the Systematic Mapping Study (SMS) is a method used to provide an overview of a broad field of interest \citep{petersen2015guidelines}. It helps to identify gaps in research, trends, and the type and quantity of evidence available, usually by quantitatively summarising and visualising a body of literature. The main aim of an SMS is to provide an overview and create a 'map' of the research done in a particular field. SMSs do not usually delve into the relationships between study characteristics and outcomes, while SQLRs go further by quantitatively analysing these characteristics and their outcomes.

Finally, an emerging trend is in automating such systematic approaches, wholly or in part, with notable examples being the Computational Literature Review \citep{mortenson2016computational} and the Smart Literature Review \citep{asmussen2019smart}. However, these are designed to automate performing SLR-like reviews on substantial bodies of literature by which a traditional human-performed review would be infeasible, rather than extracting specific features or entities from a corpus. Typically these use methods such as topic modelling \citep{blei2012probabilistic} and bibliographic analyses \citep{de2009bibliometrics}.

\subsection{Summary}
While there exist various systematic approaches to discovering knowledge in literature, our research question focuses on extracting knowledge about data entities relevant to the SDG. We propose that this knowledge about data entities, which would typically be discoverable by using data infrastructures, can be extracted from the literature systematically. We describe our method for achieving this in the next section.

\section{Methodology}
To address the research question, we develop a methodology to discover what SDG data are available based on the analysis of published literature. We based our approach on SMSs, which aim to provide an overview of a particular research area by classifying research studies to understand the covered topics and contributions of these studies \citep{petersen2015guidelines}. To scale the approach, we propose a computational pipeline based on rule-based entity extraction to analyse larger corpora to find data sources. Entity extraction is a simple but powerful technique for finding specific entities in text \citep{nadeau2007survey}, whereby one can define a set of rules to extract data entities.

SMSs have been used to map topics relating to IT; for example, where Petersen et al. \cite{petersen2015guidelines} investigated the field of software engineering, and Pedriera et al. \cite{pedreira2015gamification} explored how gamification has been applied to software engineering. Both of these studies mentioned above adopted similar research processes to obtain the mappings by using four main steps:

\begin{enumerate}
    \item Define the keywords for searching for literature from selected bibliographic databases.
    \item Filter out the results according to a defined selection and quality criteria considering the study field, language, accessibility, etc.
    \item Qualitatively analyse the remaining papers by reading them and extracting necessary features.
    \item Validate the process by evaluating the final mapping that is produced.
\end{enumerate}

Our methodology is inspired by this SMS process, where we aim to identify data entities as reported in SDG-related research literature, which are our features of interest, and create a descriptive mapping of those results. The value in taking this approach is that we can delimit the literature searches to relate to specific SDGs within the SDG framework -- that is, by each of the 17 constituent SDGs or even to a subset of the 169 individual SDG Targets -- and then discover the relevant data entities that may have nuances in different areas of sustainable development.

In the following subsections, we describe each of the following broad steps:

\begin{enumerate}
    \item Construct a corpus of literature within an SDG or SDG Target of interest (steps M1 and M2).
    \item Manually analyse the literature to extract data entities (steps M3 and M4), combining similar entities to map into higher-level data categories.
    \item Develop entity extraction rules based on the data entities found in (2) to bootstrap an automated pipeline.
    \item Construct an expanded corpus of literature within an SDG or SDG Target of interest.
    \item Apply automated pipeline (steps A1-A4) to produce the final mapping.
\end{enumerate}

\subsection{Corpus construction}
First, we must decide on the focus of our data entity discovery by choosing a particular SDG or SDG Target to constrain the document search. While it may be desirable to try and perform our data entity extraction for all SDGs collectively, we assume that a more valuable structure to the SDG data mapping is to place them within the constraints of each of the 17 SDGs or 169 SDG Targets. This places a practical constraint to make the analysis more feasible by limiting the scope of possible documents to analyse. One can then decide on a literature source (or sources) on which to perform the document search, as one would in a regular SMS, for example, using one or more bibliographic search engines such as the ACM Digital Library, Scopus or Web of Science. Two corpora are constructed - a first one that is manually analysed to develop the SDG data taxonomy, and a second larger corpus to apply the data taxonomy in the automated entity extraction pipeline to develop the final SDG data mapping.

\subsubsection{SDG-related keyword search}
Next, we must define the search with specific keywords relating to the SDG of interest. There are no standard method for this kind of search, as each SDG can be characterised with keywords in multiple ways. In one approach in a study by Brugman et al. \cite{brugmann2019expanding}, the authors identified sustainability-related courses and explored ways to enhance students' involvement in these courses. To build an inventory of sustainability-related courses, a list of keywords was defined for 17 SDGs by analysing the contents of each SDG and any possible topics. For example, where SDG 7 is about affordable and clean energy, the keywords might include "energy", "renewable", "wind", "solar", "geothermal", and "hydroelectric".

Since 2017, he Aurora Universities Network in collaboration with Scopus provides comprehensive peer-reviewed Scopus queries for each of the SDGs to search for SDG-relevant documents \citep{vanderfeesten2020search}. However, each SDG query returns a vast number of papers, in the orders of $10 \times 10^3$ for some SDGs up to $10 \times 10^6$ results with others, making it challenging to construct and analyse the corresponding corpora.

A highly restrictive strategy to define the search is to search for mentions of the relevant SDG; for example, with SDG 7, we would search for documents containing the keywords "SDG7" and "SDG 7". This has the advantage of focusing the search specifically on documents that mention the SDG of interest, ensuring a very high relevance in the search results. On the other hand, this strategy inevitably leads to missing documents of relevance to the SDG area that are not explicit in their relevance to the SDG.

\begin{table}[ht]
\vspace{12pt}
\centering
\begin{tabular}{l l} 
\hline
Search Strategy \\ \hline
Source database & - Scopus \\
Search String & - ((SDG 7) OR SDG7) AND data \\
Search Field  & - Article Title \\
              & - Abstract \\
              & - Keywords \\
Document Type & - Article \\
              & - Conference Paper \\
Source Type & - Journal \\
            & - Conference Proceeding \\
Publication Stage & - Final \\
Publication Year & - 2021, 2020, 2019 \\
                 & - 2018, 2017, 2016 \\
Language & - English \\
\hline
\end{tabular}
\vspace{0.3cm}
\caption{Summary of the literature search strategy, exemplified with the highly restrictive strategy on SDG 7.}
\label{table:strategy}
\end{table}

\subsubsection{SDG-document selection}
After retrieving a list of documents via the search, we apply any inclusion and exclusion criteria to filter the initial search results further. Regarding the inclusion criteria, firstly, the document's full text must be accessible because the text is needed to extract what data was used or referenced in the document. Not all documents are open access, and many journals and conference proceedings require paid subscriptions to access. Therefore, this criterion is highly dependent on the resources available to the researcher. The second criterion is that the paper should be SDG-related, which is initially checked manually by reading the document's title and abstract. For example, if we used the abbreviation "SDG" in the initial search, there is no guarantee that each search result is about the SDG. SDG may sometimes refer to other terms than expected. It could also be that a document mentions the relevant SDG but has a different focus. In such cases, the papers would be excluded for further analysis. A third criterion is only to include papers published after 2015. This is because the SDGs were put forward in 2015, so we only consider something before 2016 to ensure unrelated documents are excluded. Further criteria that could be applied include document types (e.g., conference or journal paper), language, and publication stage (e.g., accepted, published, pre-print etc.).

\begin{figure}[ht]
    \centering
    \includegraphics[width=\linewidth]{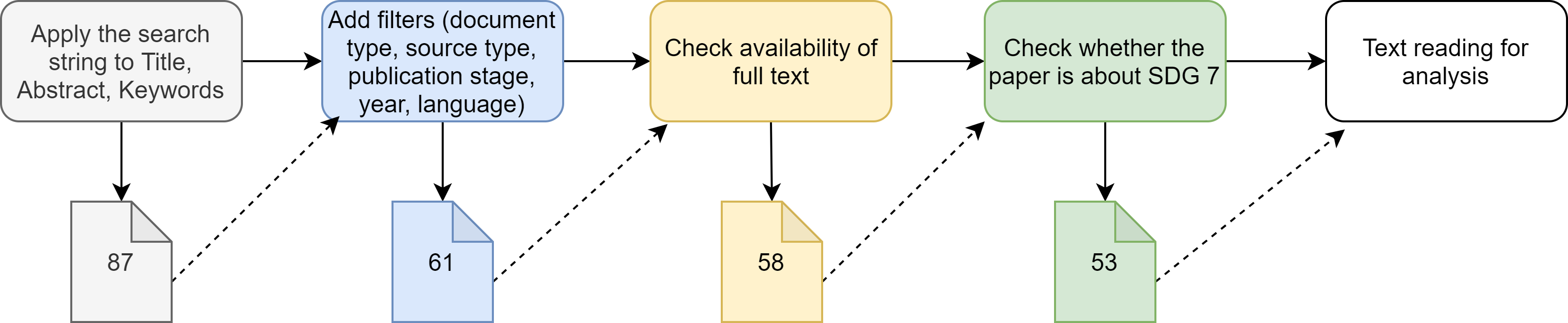}
    \caption{Illustration of the filtering applied using the inclusion and exclusion criteria, exemplified with the search results on SDG 7 the search described in Table 1.}
    \label{fig:search-results}
\end{figure}

\subsection{Corpus analysis for data source extraction}
After setting the search strategy and document selection criteria, and consequently compiling a corpus of documents to analyse, we perform a two-stage process to extract the data entities reported as being used in the published research. The first stage consists of manual qualitative coding on a literature sample to inductively build a taxonomy of data entity terms and their groupings. This is done through open coding of the literature followed by axial coding \citep{hadley2017grounded} to develop the SDG data taxonomy. This taxonomy is then used to bootstrap a set of rules implemented in Python to label documents according to the taxonomy automatically. These rules are used in an automated entity extraction pipeline to label a more extensive corpus with the data entities relating to the same SDG to develop the final SDG data mapping.

\emph{Bootstrapping} refers to a semi-supervised learning approach \citep{riloff2003manual} where a small set of manually annotated data (in our case the manually coded literature) is used to train an initial model (the resulting data taxonomy from which we develop rules), which is then used to annotate a larger set of data (the application of the rules to the expanded dataset). 

\begin{figure}[ht]
    \centering
    \includegraphics[width=\linewidth]{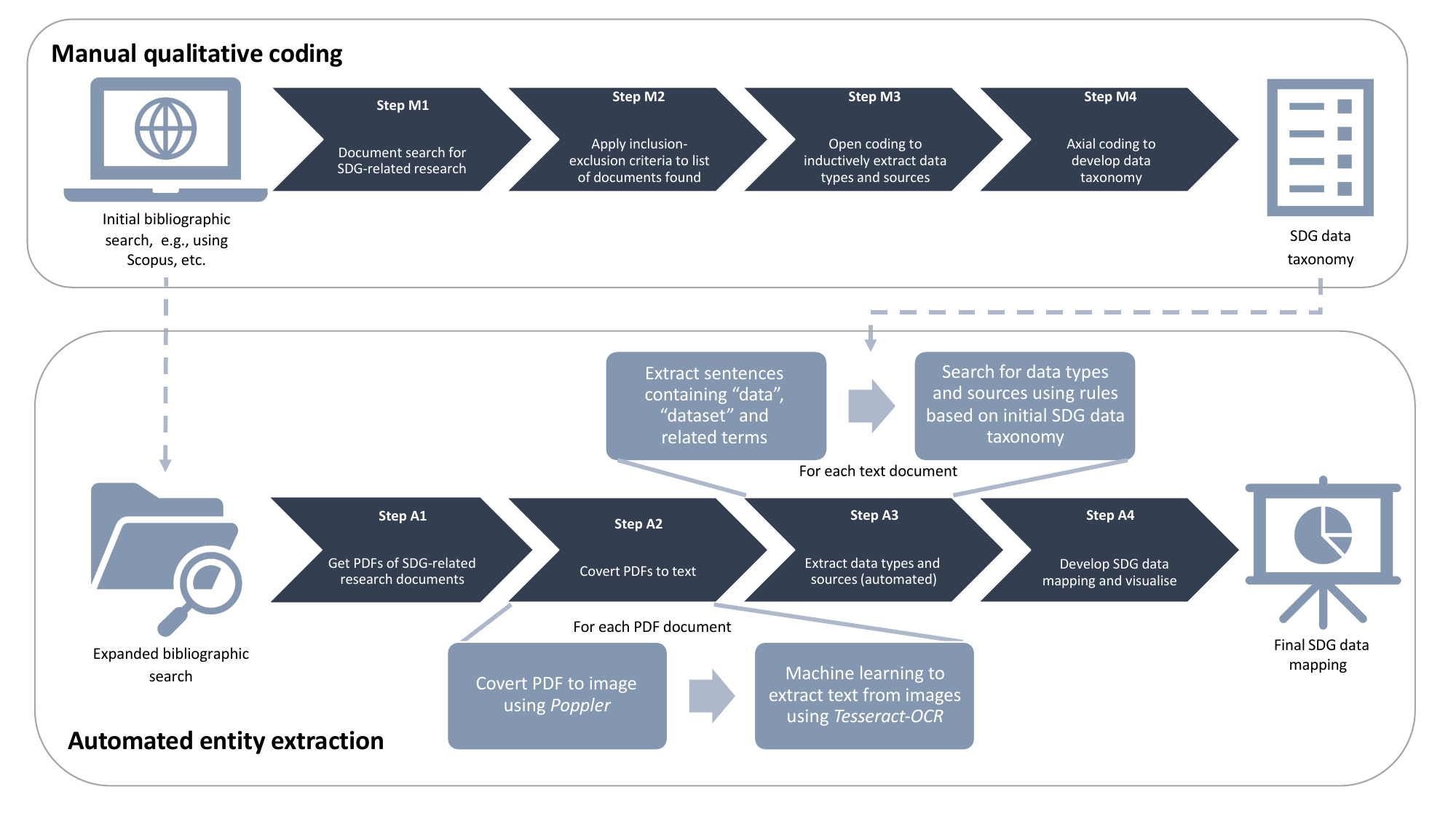}
    \caption{An overview of the qualitative coding process applied to an initial corpus of documents relating to a specific SDG or SDG Target. Note that manual coding first develops an initial SDG data taxonomy that is then used as the basis for rules developed for the automated coding stage. The automated entity extraction is then applied to an expanded document corpus to produce a final SDG data mapping.}
    \label{fig:sdg-data-process}
\end{figure}

\subsubsection{Manual Qualitative Coding of Data Sources}
The first stage of the analysis consists of a manual qualitative coding process divided into four steps, as shown in Figure \ref{fig:sdg-data-process} in steps M1-M4. Steps M1 and M2 implement the search strategy and apply the inclusion/exclusion criteria discussed above. Step M3 requires the researcher to read each document to understand the data used or referenced by the document. The data described is usually collected by the document authors or sourced from other organisations or public databases. No matter where the data is from, it should be recorded because any data could be a potential for the SDG of interest. Traditional data sources like interviews and surveys may be mentioned, as well as sources from databases or other digital systems. For some documents, the authors might not describe the data source but illustrate what the data relates to (e.g., electricity) or the data type (e.g., hourly household smart electricity meter readings). Therefore, data types should be considered as another type of data entity. A key consideration during this process is that the researcher should inductively develop the categorisation without any preconceived expectation of the resulting taxonomy. Finally, in step M4, these results are summarised as a taxonomy and labelling of the papers.

\subsubsection{Automated Entity Extraction of Data Sources}
The second stage of the analysis consists of an automated entity extraction process divided into four steps, as shown in Figure \ref{fig:sdg-data-process} in steps A1-A4, implemented as a data processing pipeline in Python. The main idea behind this stage is to take the SDG data taxonomy developed from the manual qualitative coding by the human researcher and implement entity extraction rules as a decision list \cite{rivest1987learning}. We can then run an automated analysis on an expanded document corpus to label each document with data entity reported. Step A1 involves getting PDF files of each document from the list of documents determined in the initial stage. Next, the PDFs are converted into raw text in step A2. This is not trivial as PDF documents vary in formatting and layouts, so our pipeline transforms the pages of each PDF into images using the Poppler software package, where the text is then extracted using the Tesseract-OCR (Optical Character Recognition) engine to text files. Not only does this help us more comprehensively capture the text of a document, but it also captures text embedded in figures such as charts and images that may contain text referencing data. In step A3, our pipeline then performs some data cleaning and finds sentences that mention variations on "data" or "data set" and applies the decision list rules to label each document with a data entity. Finally, in step A4, we produce the final mapping as a table of documents and labels and plot the corresponding visualisations of the distributions as sunburst charts \cite{woodburn2019interactive} using the Plotly Python package.

\section{Exemplar analysis on SDG 7}
We applied the methodology described in the previous section to analyse literature drawn from SDG 7 (Affordable and Clean Energy) to demonstrate our approach to discovering SDG data.

\subsection{Constructing an SDG 7 Data Taxonomy}

Scopus was used as the bibliographic database, and the keywords "SDG 7", "SDG7", and "data" were applied to the title, abstract, and keywords search fields, resulting in 87 candidate documents for inclusion. We then applied each selection criteria to focus the document corpus further to arrive at a final corpus of 53 documents to use in the first stage of our methodology, manual quantitative coding. A summary of the document search and selection that constitute steps M1 and M2 are shown earlier in Table \ref{table:strategy}, and the effect of applying each selection criterion is shown in Figure \ref{fig:search-results}.

The final list of documents was exported from the Scopus database to an Excel spreadsheet containing the document title, abstract, keywords, and Web links to the Scopus record. The researcher then read each document online while performing open coding and added a short summary description about each in the spreadsheet, along with any data entity of interest noted (step M3). Once each document had been analysed, the open codes were reviewed via axial coding and grouped into higher-level classifications thus building up an initial SDG 7 data taxonomy (step M4).

This taxonomy organised data entities of interest into two broad groups: data sources and data types. The documents were then further organised into hierarchical classification levels based on shared traits to better understand the taxonomy.

\subsubsection{Data sources}
The data sources were initially divided into traditional statistics and organisational databases. Traditional statistics included data collected from surveys, censuses, interviews, focus groups, and questionnaires. Organisational databases provided an easier way for researchers to find where to obtain the needed data. Under this category, two subgroups were put forward according to the scope of an organisation or a database so that it was either international or national.

As the largest international organisation in the world, the United Nations serves to maintain world peace, keep friendly relations among countries, and provide better lives for human beings. Apart from the UN statistical databases, four more agencies of the UN were referred to as data sources in the papers. The Food and Agriculture Organisation (FAO) cares about ending hunger and food security issues; the Sustainable Development Solutions Network (SDSN) seeks to better solutions for achieving SDGs through cooperation among worldwide experts; the United Nations Educational, Scientific and Cultural Organisation (UNESCO) is responsible for the world peace in education, science, and culture; the World Health Organisation (WHO) works to promote public health worldwide. 

Four European Union (EU) organisations were also components of international organisations. Eurostat provides statistics on various topics covering different social themes, like energy, agriculture, and more for the member states. The Emissions Database for Global Atmospheric Research (EDGAR) records statistics for emissions of greenhouse gases like carbon dioxide on Earth. As an EU program focusing on our environment, Copernicus provides data and services concerning the atmosphere, marine, land, climate change, and more. European Soil Data Center (ESDAC) delivers soil-related data at different European scales. 

As for those international organisations that neither belongs to the UN nor the EU were categorised into the same group. The World Bank is a global partnership aiming to end world poverty and reach shared prosperity, containing indicators as measurements and series data on diverse topics like energy. International Energy Agency (IEA) is a platform for delivering comprehensive global energy statistics of high quality, including indicators, energy demand, prices, and more. Apart from the World Bank and IEA, other organisations under this category pay attention to energy and economics, country spatial data, emissions, and population. 

\subsubsection{Data types}
The data types were divided into resource, weather, geographic, and sensor data. Different kinds of resource data were analysed in the obtained papers. Renewable energy was one of the core contents of SDG 7. Biomass energy, solar energy, and heat, typical examples of renewable energy, were discussed and referred to in the articles. Mineral data was also mentioned because it met some requirements for technologies to generate clean energy, like magnets in wind turbines. Electricity data was another element under the category of resource data. An indicator of SDG 7 is the "proportion of population with access to electricity", so electricity data has a close connection with SDG 7, and it needs to be considered when measuring the progress of SDG 7. Energy use could also affect other resources, so water and land use data were discussed in the selected papers. As for the weather data, it was also mentioned in the papers. The weather affected the generation of wind, water, and bioenergy. The geographic data included satellite imagery, Geographic Information System (GIS) data, Global Positioning System (GPS) data, and OpenStreetMap data. These kinds of data helped to understand the distribution of energy consumption. The data collected from sensors was another typical data type. As a critical component of IoT, the sensor could be applied in various fields. For example, it could collect energy consumption and waste to measure energy efficiency. All the data types described above were obtained manually from the selected 53 papers.

\subsection{Data Entity Extraction for SDG 7}
Next, we use the SDG 7 data taxonomy from the first stage to prepare for the second stage of analysis, automated entity extraction, by developing a decision list that classifies each document entry based on the SDG 7 data taxonomy from the first stage. For example, in the manual qualitative coding stage it was found that the International Energy Agency was a relevant data entity (and data source), and this was identified by noting the words "International Energy Agency" and its abbreviation "IEA" within the context of a sentence about data or data sets. Based on this observation from the open coding, we model this classification computationally as a Boolean decision function in Python code. We build the decision list with similar functions for each code developed in the open and axial coding steps. The first step, A1, in the entity extraction pipeline, is to obtain the PDF file for each of the documents. This was done by manually downloading the PDFs from the relevant publisher and storing them locally in a folder on a local disk. Next, steps A2 to A4 are run where Python code loads each document from disk and converts the PDFs to text (step A2), applies the decision list classifier (step A3), and compiles the final mapping and corresponding visualisation (step A4). 

\subsection{Data discovered for SDG 7}
\begin{figure}[ht]
    \centering
    \includegraphics[width=\linewidth]{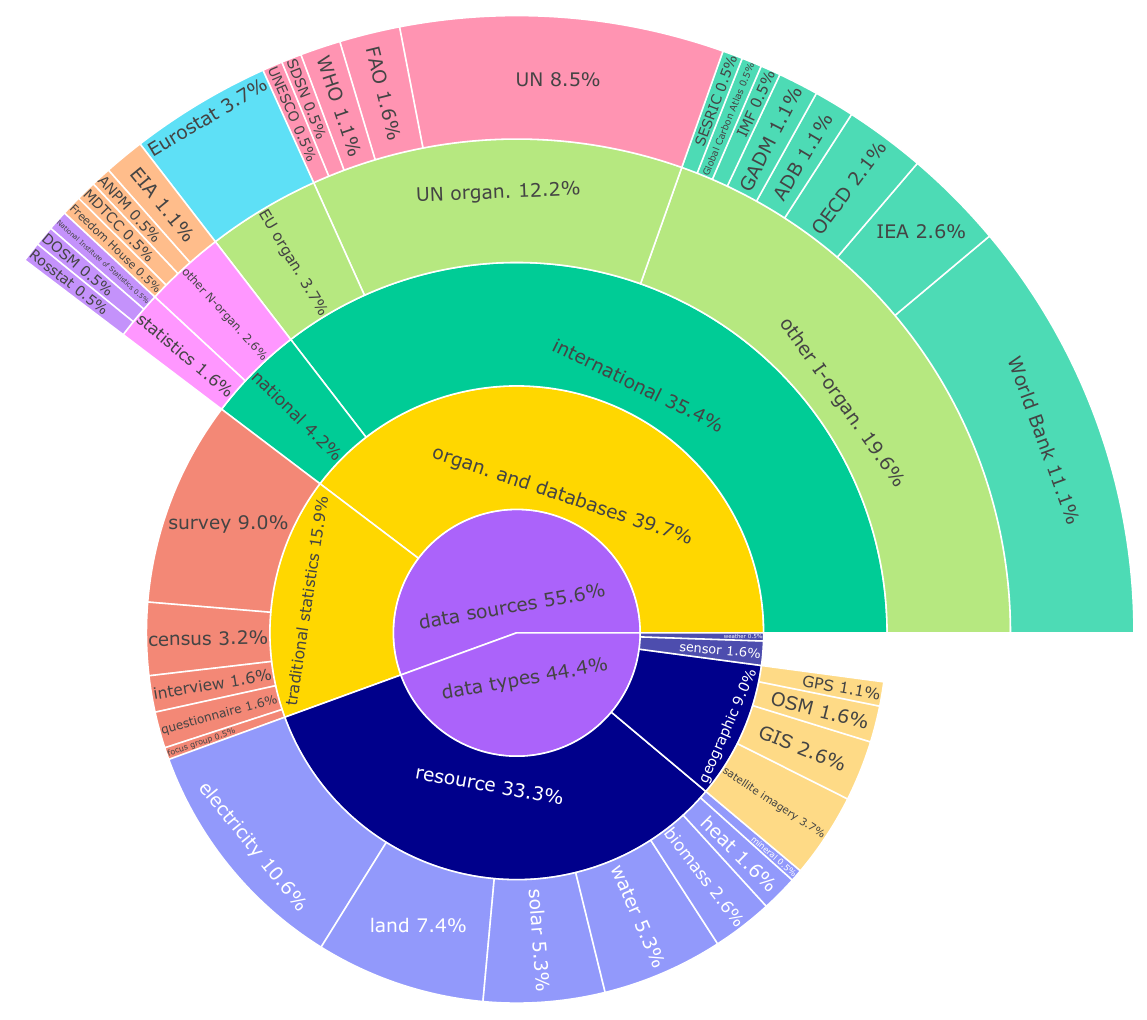}
    \caption{An example sunburst chart showing the results of an application of the entity extraction approach for SDG 7. The centre of the chart shows the highest level of the taxonomy hierarchies, dividing the categories found as data sources and data types, followed by more specific subcategories, down to their lowest-level data entities.}
    \label{fig:c-pie}
\end{figure}

In the results obtained from the automated entity extraction, the percentage of data sources and data types accounted for was nearly half, 55\%, and 45\%. A visualisation of the summary distribution of data entities and categories for the SDG 7 data mapping is shown in Figure \ref{fig:c-pie}

In the category of data sources, the number of organisations and databases that were referred to was more than twice that of traditional statistics. The survey was still the most popular conventional way of data collection among these studies. International organisations and databases were the traditional data sources, accounting for more than one-third of the total data entities. The World Bank, databases of the UN, and Eurostat were the top three frequently referred data sources, meaning they were the primary sources among the SDG 7 related papers.

Regarding the data types, the source and geographic data were discussed more often than the sensor data and weather. The proportion of diverse resource and geographic data chosen in the research was around 95\% of data types. Data on electricity, land use, solar, and water use were the most popular being used to monitor the progress of SDG 7.

\section{Discussion}

This proposed approach augments the traditional systematic review with automation, which is crucial for several reasons. Firstly, it effectively fills a gap in existing literature review methodologies by demonstrating how a systematic review or mapping process can focus on specific features, on the data types and sources that are reported in SDG research literature. Exposing the nature of these data in SDG literature is also a means to discovery on two fronts: understanding the current state of data usage for SDGs; and potentially uncovering new or scarcely used data types and sources. The discovery of innovative data uses for the SDGs would push the boundaries of future SDG research and encourage diversity in data innovation for sustainable development.

Above all, automation is critical in this approach. By transferring the burden of manual entity extraction to an automated pipeline, researchers can invest their time and resources in more profound analysis and interpretation. The manual qualitative coding on a smaller sample of SDG literature first builds the taxonomy specific to that literature. Then, the implementation of the automated stage in the data entity extraction process brings some specific advantages:

\begin{enumerate}
    \item As a computational pipeline, the automated application of the rules facilitates more scalable processing of a large volume of documents, overcoming the time-consuming and labour-intensive constraints of manual coding.
    \item Automation ensures consistent application of rules across the documents, avoiding inconsistencies that might arise in manual analysis. This consistency promotes reproducibility enabling other researchers to validate and replicate the process and the results \citep{peng2011reproducible}.
    \item A pipeline such as this can be periodically run to facilitate longitudinal analysis that reveals trends in trends in using certain data over time.
    \item Automating this process also allows for more effective resource allocation; human expertise can be redirected to perform tasks that require deeper interpretation and strategic decision-making.
\end{enumerate}

It should be noted that while automation can accelerate the data extraction process, it should not be viewed as a complete replacement for manual coding. Despite the benefits outlined above, automated techniques such as this may need more nuanced understanding and judgment than humans, especially in complex, context-dependent analysis. Human-led systematic mapping studies are generally meticulous and less prone to errors when compared to automated processes. An automated process based on rules developed from a small sample may not accurately reflect the content of an expanded corpus, potentially leading to omissions or misinterpretations in the data. Nonetheless, we submit that our approach provides a valuable methodological contribution to using IT to support sustainable development research, and a methodological approach to understanding how data can be used to make progress towards the SDGs.

\section{Conclusion and future work}
In this paper, we presented an approach to discovering data of interest to the SDGs, combining the SMS approach used to bootstrap an automated pipeline for data entity extraction. The systematic literature search of data relating to a specific SDG or SDG Target and subsequent manual analysis through open and axial coding are used to construct a data taxonomy on a sampled base of the literature of interest. The data taxonomy is then used as the basis for a set of rules implemented in a computational pipeline to process an expanded corpus.

While this approach contributes a method to data discovery employing a literature-based approach, it has potential drawbacks, which we have discussed in the previous section. Future work on this approach could focus on addressing the drawbacks discussed. A significant step could be refining the automated entity extraction process to handle complex and nuanced data entities more effectively. This could be achieved through more sophisticated NLP or machine learning techniques.

Another promising direction is to improve the handling of non-traditional or emerging data sources reported in SDG literature. The current mapping only provides a quantitative summary distribution of data entities. However, it may be interesting to investigate how one can quantify novelty, such as in \citep{kyriakou2022novelty}, of data uses in SDG research to determine innovative uses of data. Finally, it may also be interesting to see researchers apply our method across the range of SDGs to review the landscape of data used in sustainable development research.


\bibliographystyle{ACM-Reference-Format}
\bibliography{sample-base}

\appendix

\section{Data Availability}

The source code and data for this paper can be found in: \url{https://github.com/UppsalaIM/Data_Discovery_for_the_SDGs/}. Operating system: Platform independent. Programming language: Python. Other requirements: Poppler and Tesseract-OCR. License: MIT license for all code written by the authors.

\end{document}